\documentstyle[12pt]{article}
\begin{document}
\rightline{TUHE9561}
\begin{center}
{\bf\large Sum rules for charmed baryon masses}\\
\vspace{16pt}
Jerrold Franklin\\
{\it Department of Physics,Temple University,\\
Philadelphia, Pennsylvania 19022}\\
June 1995
\end{center}
\begin{abstract}

The measured masses of the three charge states of the charmed $\Sigma_c$ baryon
are found to be in disagreement with a sum rule based on the quark model, but
relying on no detailed assumptions about the form of the interaction.  This
poses a significant problem for the charmed baryon sector of the quark model.
Other relations among charmed baryon masses are also discussed.

\end{abstract}
\vspace{.5in}

In recent years, measurements have been made\cite{pdg} of the masses
of the three charge states of the charmed $\Sigma_c$ baryon.
These measurements can be applied to sum rules\cite{cb}
that were derived some time ago using fairly minimal assumptions
within the quark model.  The sum rules depend on standard quark
model assumptions, and the additional assumption that the interaction
energy of a pair of quarks in a particular spin state does
not depend on which
baryon the pair of quarks is in.  No assumptions are made about the
type of potential, and no internal symmetry is assumed.

The $\Sigma$ sum rule relates  electromagnetic mass differences
of the $\Sigma_c$ baryon with corresponding mass differences of the  $\Sigma$
and $\Sigma^*$\cite{cb}
\begin{eqnarray}
D_{uu}+D_{dd}-2D_{ud}&=&
\Sigma^++\Sigma^- -2\Sigma^0=1.7\pm 0.2\\
&=&\Sigma^{*+}+\Sigma^{*-} -2\Sigma^{*0}=2.6\pm 2.1\\
&=& \Sigma_c^{++}+\Sigma_c^0 -2\Sigma_c^+=-2.1 \pm 1.3.
\end{eqnarray}
The baryon symbol has been used as its mass, and the
$D_{ij}$ represent the two body interaction energies between pairs of
quarks in states of spin one.
The sum rule relating the $\Sigma_c$ and the $\Sigma$, which is among the most
rigorous in I, is violated by three
standard deviations.

 Although no assumption has been made about the form of the interaction
energies, this sum rule is probably purely electromagnetic because the QCD mass
corrections
to the combination $D_{uu}+D_{dd}-2D_{ud}$ cancel to first
order in the ratio $\delta=(m_d -m_u)/m$, (m is the average of the nucleon
quark
masses.)
and the second order correction is negligible.
The equality represented by the sum rule follows because the
two body interaction energies
given by the $D_{ij}$ are the same for each combination of baryons.
This is because they are all in the same spin one state
for corresponding pairs of quarks.  A number of
two body interaction energies (also involving other spin states)
cancel  in the
linear combinations formed in the sum rule.

It has been suggested that there should
be some dependence of the two body interaction energy on the third quark in
the baryon.\cite{db1,db2}  We have  estimated this effect,
following the procedure suggested in Ref. \cite{db2} using their
parameters. The net change in the $\Sigma_c$ sum is only 0.1 MeV. so that
the sum rule seems to be quite robust with respect to this type of correction.
One reason for this is that all cancellations of interaction terms take place
between pairs of quarks that are in corresponding positions in the baryons.
Of the nine original interaction terms in each combination of
$\Sigma$ baryons, the six that cancel are
essentially unaffected by this type of mass correction because of
the cancellation of mass effects to first order.

In looking more deeply at the $\Sigma-\Sigma_c$ sum rule, the
+1.7 MeV for the uncharmed $\Sigma$ combination seems to be sensible,
but the -2.1 MeV for the $\Sigma_c$ is difficult to understand.  If
it is purely electromagnetic,
the mass difference for the $\Sigma$s  is given by\cite{sqm}
\begin{equation}
D_{uu}+D_{dd}-2D_{ud}=\alpha_{em}<1/r> - D_m,
\label{eq:em}
\end{equation}
where r is the distance between the two nucleon quarks.
The magnetic contribution is given by
\begin{equation}
D_m=\frac{2\pi\alpha_{em}}{3m^2}|\psi(0)|^2].
\end{equation}
The magnetic contribution can be estimated by comparing it to a corresponding
QCD contribution\cite{sak,fl,db4}
\begin{equation}
D_{QCD}=[2(\Sigma^{*0}-\Sigma^0) + 3(\Sigma^0 - \Lambda^0)]/12=51 MeV.
\end{equation}
Then
\begin{equation}
 D_m=\frac{3\alpha_{em}}{2\alpha_{QCD}}D_{QCD}=1.0 MeV,
\label{eq:lm}
\end{equation}
where we have used\cite{fl'} $\alpha_{QCD}=0.56$.
Using this value of $D_m$ in Eq. (\ref{eq:lm}) results in
\begin{equation}
<1/r> = 1/0.53 fm \hspace{.5in}and\hspace{.5in}|\psi(0)|^2=1/(1.0 fm)^3.
\end{equation}
These values are reasonable ones for the expected baryon size.
On the other hand, even the sign of the $\Sigma_c$ sum is hard to understand.
It is difficult to think of any quark wave function and masses
that could lead to a negative sign for Eq. (\ref{eq:em}).  If future
experiments
do not result in a different value for the combination
$\Sigma_c^{++}+\Sigma_c^0 -2\Sigma_c^+$, the quark model for charmed baryons
would require considerable revision.  That is the main conclusion of this
paper.

Other sum rules given in I can be applied to
measurements of the masses of the $\Omega_c^0$  and
the two charge states of the $\Xi_c$.  We present these here, but with the
caveat that they would not apply if the above violation of the more rigorous
electromagnetic sum rule for the charmed $\Sigma_c$ baryons cannot be resolved.
The first of these is\cite{cb}
\begin{eqnarray}
D_{uu}+D_{ss}-2D_{us} & = & \Delta^{++}+\Xi^{*0}-2\Sigma^{*+}=-3\pm 1\\
 & =&  \Sigma_c^{++}+\Omega_c^0-2\Xi_c^{\prime +}.
\end{eqnarray}
We use the prime on $\Xi_c^{\prime +}$ to signify that its u
and s quarks are in a spin one state.  The unprimed $\Xi_c^+$
has the u and s quarks in a spin zero state.
Note that this convention is opposite to the notation in I.

We can use this sum rule  to predict
the mass of the $\Xi_c^{\prime +}$  to be
\begin{equation}
\Xi_c^{\prime +}=2583\pm 3.
\end{equation}
This is consistent with the prediction $\Xi_c^{\prime +}=2580\pm 20$
in Ref. \cite{db3}.

  If we modify this sum rule by the mass corrections
of Ref. \cite{db2},
we find that individual terms (there are 18 in the sum rule)
can be changed by as much as 5 MeV
in substituting a c quark for a spectator u quark.
However, these changes
tend to cancel out in taking the mass differences,
and the net contribution of these effects on the sum rule would be
to raise the predicted mass of the $\Xi_c^{\prime +}$ by only 5 MeV.
Incidentally, the sum rule makes it clear that the observed
charmed $\Xi$ is the $\Xi_c^+$, since the $\Xi_c^{\prime +}$
would violate the sum rule by a large amount if it had the mass
of the observed $\Xi_c^+$(2465 MeV).

A combination of sum rules from I can be used to predict the isospin breaking
mass difference of the $\Xi_c^{\prime}$ baryon
\begin{equation}
\Xi_c^{\prime 0}-\Xi_c^{\prime +}=(\Xi^{*-}-\Xi^{*0})
-(\Sigma^{*0}-\Sigma^{*+})+(\Sigma_c^+-\Sigma_c^{++})=3.0\pm 1.4.
\end{equation}
The interaction energy difference in Eq.(12) comes from the QCD $1/m_im_j$
interaction as well as electric Coulomb and magnetic dipole-dipole
interactions,
similar to those in Eq. (4).

However, this prediction is made ambiguous by the experimental failure
of the $\Sigma_c$ sum rule.  There are theoretically equivalent expressions
for the $\Xi_c^{\prime}$ mass difference given by
\begin{eqnarray}
\Xi_c^{\prime 0}-\Xi_c^{\prime +}&=&(\Xi^{*-}-\Xi^{*0})
-(\Sigma^{*-}-\Sigma^{*0})+(\Sigma_c^0-\Sigma_c^{+})\nonumber\\
&=&-1.7\pm 1.0,\\
\Xi_c^{\prime 0}-\Xi_c^{\prime +}&=&(\Xi^{*-}-\Xi^{*0})
-\frac{1}{2}[(\Sigma^{*0}-\Sigma^{*++})+(\Sigma_c^0-\Sigma_c^{++})]
\nonumber\\
&=&0.6\pm 0.8.
\end{eqnarray}
The inconsistency of these theoretically equivalent predictions
 highlights the failure of the $\Sigma_c$  sum rule.


\begin{thebibliography}{8}
\bibitem{pdg} Review of Particle Properties, Physical Review
	{\bf D50}, 1173 (1994).\\
	All baryon masses (in MeV) are taken from this reference.
\bibitem{cb} J. Franklin, Phys. Rev. {\bf D12}, 2077 (1975).  We refer to this
paper as I.
\bibitem{db1} Y. Wong and D. B. Lichtenberg, Phys. Rev. {\bf D42}, 2404 (1990).
\bibitem{db2} R. Roncaglia, A. Dzierba, D. B. Lichtenberg, and E. Predazzi,
              Phys. Rev. {\bf D 51}, 1248 (1995).
\bibitem{sqm} J. Franklin Phys. Rev. {\bf 172}, 1807 (1968).
\bibitem{sak} A. D. Sakharov, Zh. Eksp. Teor. Fiz. {\bf 79}, 350 (1980)\\
	(Sov. Phy. JETP {\bf 52}, 175 (1980)).
\bibitem{fl} J. Franklin and D. B. Lichtenberg, Phys. Rev. {\bf D25},
	1997 (1982).
\bibitem{db4} M. Anselmino, D. B. Lichtenberg, and  E. Predazzi,
	Z. Phys. {\bf C48},605 (1990).
\bibitem{fl'} This value  is slightly different	than given
		 in Ref. \cite{fl} because the measured
		$\Sigma^--\Sigma^+$ mass difference has changed slightly.
 Using $\alpha_{QCD}=0.39$, as suggested in Ref. \cite{db2}
              would give $D_m =1.4$ MeV.
\bibitem{db3} R. Roncaglia, D. B. Lichtenberg, and E. Predazzi, Indiana
	University preprint IUHET 293 (1995), Phys. Rev. D
	(to be published Aug. 1, 1995).
\end{thebibliography}
\end{document}